\documentclass[amsmath,amssymb]{nature}

\usepackage{ifthen}		
\newboolean{SubmittedVersion}
\setboolean{SubmittedVersion}{False} 	

\usepackage{graphicx}
\usepackage{amsmath}
\usepackage{amssymb}
\DeclareMathAlphabet{\mathpzc}{OT1}{pzc}{m}{it}

\begin{document}

\title{Realization of Two-Dimensional Spin-orbit Coupling for Bose-Einstein Condensates}

\author{Zhan Wu$^{1,2}$, Long Zhang$^{1,3}$, Wei Sun$^{1,2}$, Xiao-Tian Xu$^{1,2}$, Bao-Zong Wang$^{1,3}$, Si-Cong Ji$^{1,2}$, Youjin Deng$^{1,2}$, Shuai Chen$^{*1,2}$, Xiong-Jun Liu$^{*3,4}$, and Jian-Wei Pan$^{*1,2}$}

\maketitle

\begin{affiliations}
 \item Shanghai Branch, National Laboratory for Physical Sciences at Microscale and Department of Modern Physics, University of Science and Technology of China, Shanghai 201315, China
 \item Synergetic Innovation Center of Quantum Information and Quantum Physics, University of Science and Technology of China, Hefei, Anhui 230026, China
 \item International Center for Quantum Materials, School of Physics, Peking University, Beijing 100871, China
 \item Collaborative Innovation Center of Quantum Matter, Beijing 100871, China
\end{affiliations}

\begin{abstract}
Cold atoms with laser-induced spin-orbit (SO) interactions provide intriguing new platforms to explore novel quantum physics beyond natural conditions of solids. Recent experiments demonstrated the one-dimensional (1D) SO coupling for boson and fermion gases. However, realization of 2D SO interaction, a much more important task, remains very challenging. Here we propose and experimentally realize, for the first time, 2D SO coupling and topological band with $^{87}$Rb degenerate gas through a minimal optical Raman lattice scheme, without relying on phase locking or fine tuning of optical potentials. A controllable crossover between 2D and 1D SO couplings is studied, and the SO effects and nontrivial band topology are observed by measuring the atomic cloud distribution and spin texture in the momentum space. Our realization of 2D SO coupling with advantages of small heating and topological stability opens a broad avenue in cold atoms to study exotic quantum phases, including the highly-sought-after topological superfluid phases.
\end{abstract}

Spin-orbit (SO) interaction of an electron is a relativistic quantum mechanic effect, which characterizes the coupling between motion and spin of the electron when moving in an electric field. In the rest frame the electron experiences a magnetic field which is proportional to the electron velocity and couples to its spin by the magnetic dipole interaction, rendering the SO coupling. The SO interaction plays essential roles in many novel quantum states of solids. The recent outstanding examples include the topological insulators, which have been predicted and experimentally discovered in two-dimensional (2D) and 3D materials~\cite{hasan10,qi11}, and the topological superconductors~\cite{Read2000,Kitaev2001}, which host exotic zero-energy states called Majorana fermions~\cite{Wilczek2009,Franz2013} and still necessitate rigorous experimental verification. For topological insulators, the strong SO interaction leads to the so-called band inversion mechanism which drives a topological phase transition in such systems~\cite{Zhang2006,Zhang2009}. In superconductors, a triplet $p$-wave pairing is generically resulted when SO coupling is present, for which the superconductivity can be topologically nontrivial under proper conditions~\cite{Alicea2012}.

Recently, considerable interests have been drawn in emulating SO effects and topological phases with cold atoms, mostly driven by the fact that cold atoms can offer extremely clean platforms with full controllability to explore such exotic physics. In cold atoms the synthetic SO interaction can be generated by Raman coupling schemes which flip atom spins and transfer momentum simultaneously~\cite{Ruseckas2005,Osterloh2005}, and so far the 1D SO interaction~\cite{Liu2009} has been successfully demonstrated in experiment for both cold boson~\cite{Lin,Jinyi2012} and fermion degenerate gases~\cite{MIT,Wang}. With the 1D SO coupling, which corresponds to an Abelian gauge potential, a few novel effects can be studied, such as the magnetized or stripe ground states for bosons~\cite{Zhai2010,Wu2011,Ho2011,Li2012} which have been observed in experiment~\cite{Cong2014}, novel spin dynamics~\cite{Galitski2013,Jeffrey2015}, and 1D insulating topological states for fermions~\cite{Goldman2014}. However, realizing higher dimensional SO couplings, which corresponds to non-Abelian gauge potentials~\cite{Ruseckas2005,Osterloh2005}, can enable the study of much broader range of nontrivial quantum states which do not exist in systems with only 1D SO interaction. For example, the recently discovered topological insulators are driven by 2D and 3D SO interactions~\cite{hasan10,qi11}.
Further, since the long-range superfluid order cannot exist for a 1D cold atom system, to have a 2D SO interaction is the minimal requirement to reach a gapped topological superfluid phase through a conventional $s$-wave superfluid state~\cite{Chuanwei2008,Sato2009}.

The generation of 2D SO interactions in a realistic cold atom system is by no means straightforward. Thus far considerable efforts have been made in proposing different schemes for 2D and 3D SO couplings~\cite{Ruseckas2005,Chuanwei2,Ian,magnetic1,magnetic2}, whereas their experimental realizations are yet to be available for degenerate atom gases. The typical challenges in the proposals are that the schemes either reply on complicated Raman coupling configurations or suffer heating which cause instability of the realization~\cite{note}. Very recently, it was proposed that 2D SO coupling can be realized by a simple optical Raman lattice scheme which applies two pairs of light beams to create the lattice and Raman potentials simultaneously~\cite{Liu2014}. Nevertheless, this scheme contains two Raman transitions with their relative phase having to be locked, which is a difficult task since the two Raman potentials are induced by four independent lights of different frequencies. In this work, we propose a new minimal scheme to overcome all the challenges in the previous work~\cite{Liu2014} and successfully demonstrate the realization of 2D SO coupling with $^{87}$Rb Bose-Einstein condensates (BECs).

\begin{figure}[t]
\begin{center}
\ifthenelse{\boolean{SubmittedVersion}}{}{\includegraphics[width= 125 mm]{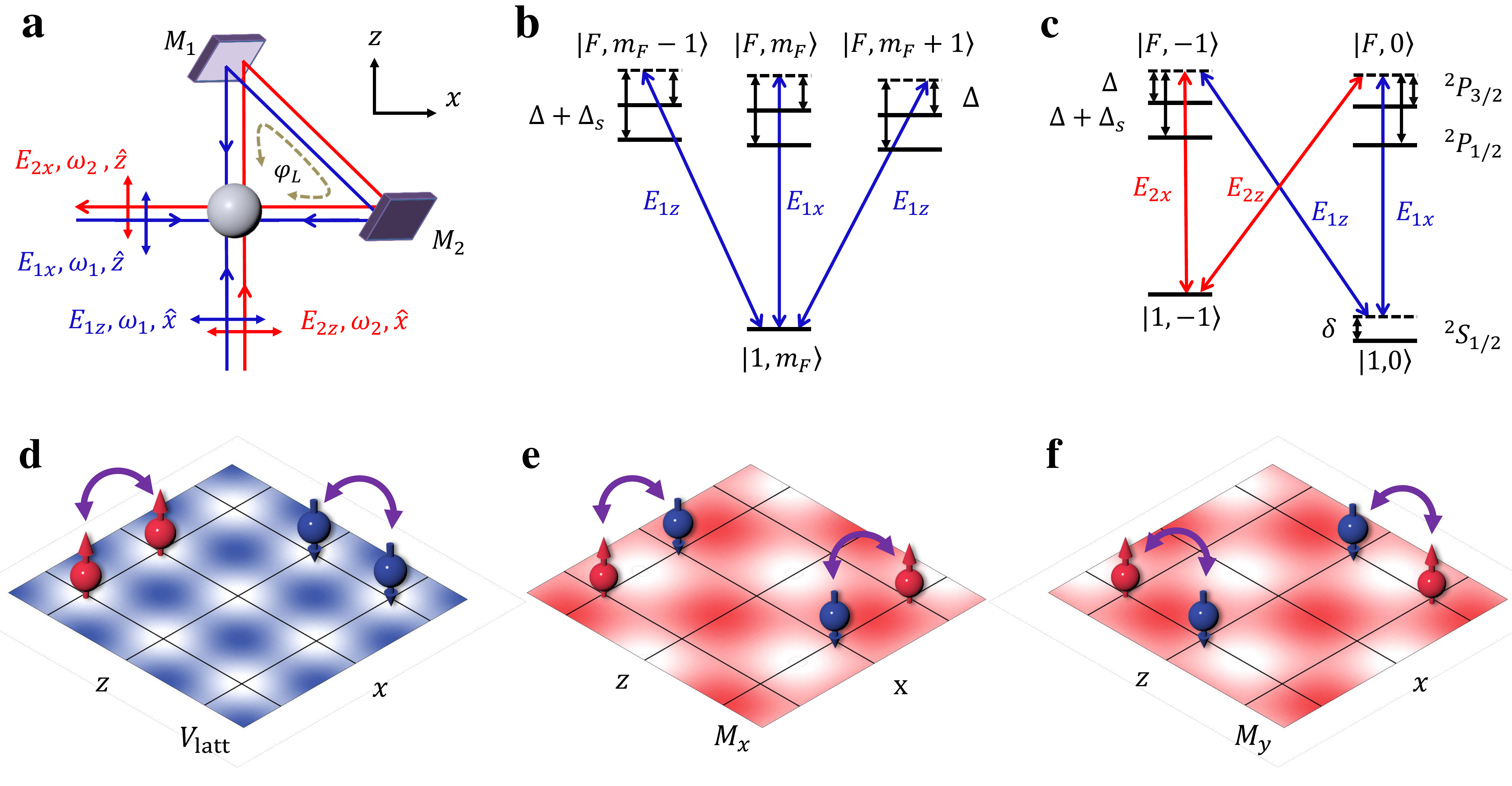}}
\end{center}
\caption{{\bf Proposal of the optical Raman lattice scheme.} ({\bf a}) Sketch of the setup for realization. ({\bf b}) Lattice potentials generated by $E_{1x,1z}$ for the states $|1,m_F\rangle$ ($m_F=0,-1$). ({\bf c}) The two periodic Raman potentials are generated through a {\it double}-$\Lambda$ type configuration of atoms. ({\bf d}) The profile of lattice potential for $V_{0x}=V_{0z}=5E_{\rm r}$. Raman potentials $M_x$ ({\bf e}) and $M_y$ ({\bf f}) for $M_{0x}=M_{0y}=1.2E_{\rm r}$. The Raman potential $M_x$ is antisymmetric (symmetric) with respect to the lattice site along $x$ (z) direction, while $M_y$ is antisymmetric (symmetric) with respect to the lattice site along $z$ (x) direction.} \label{figure1}
\end{figure}


We aim to realize 2D SO coupling and topological band for ultracold atoms on
a square lattice, of which the Hamiltonian reads
  \begin{equation}\label{Ham}
H=\left[\frac{\hbar^2{\bf k}^2}{2m}+V_{\rm latt}(x,z)\right]\otimes{\bf 1}+{\cal M}_{x}(x,z)\sigma_x+{\cal M}_{y}(x,z)\sigma_y+m_z\sigma_z \; ,
\end{equation}
where ${\bf 1}$ is the $2\times2$ unit matrix, $\sigma_{x,y,z}$ are Pauli matrices acting on the spins, $m$ is mass of an atom, $V_{\rm latt}$ denotes
the lattice potential in the $x$-$z$ plane,
${\cal M}_{x,y}$ are periodic Raman coupling potentials, and $m_z$ represents a tunable Zeeman field.
Atoms can hop between nearest-neighboring sites due to lattice potential as well as the Raman coupling terms.
Note that $V_{\rm latt}$ is {\it spin-independent} and can induce hopping which conserves the atom spin.
In contrast, the Raman-assisted hopping flips the atom spin.
The overall effect of hopping along $\hat{x}$ and $\hat{z}$ directions shall give a novel 2D SO coupling, which can lead to nontrivial topological bands for the square lattice.

Here we propose to realize Hamiltonian~(\ref{Ham}) through a minimal scheme, which is generic and applicable to both boson and fermion atoms~\cite{Long2015}. Fig.~\ref{figure1} illustrates the realization in the $^{87}$Rb Bose gas, with $|\uparrow\rangle\equiv|1,-1\rangle$ and $|\downarrow\rangle\equiv|1,0\rangle$, and the hyperfine state $|1,+1\rangle$ being removed by a sufficiently large two-photon detuning. The minimal ingredients of the realization include a blue-detuned square lattice created with two light components denoted by the blue lines, and the periodic Raman potentials generated together with additional light components denoted by the red lines [Fig.~\ref{figure1}{\bf a}]. The both ingredients can be achieved with a single in-plane ($x$-$z$) linearly polarized laser source. In particular, the optical lattice is generated by $E_{1x}$ and $E_{1z}$ (blue lines) which are incident from horizontal ($x$) and vertical ($z$) directions, respectively [Fig.~\ref{figure1}{\bf a}], and can be created from a single light of frequency $\omega_1$ by a beam splitter in the experiment. The lights are reflected by two mirrors $M_1$ and $M_2$, and form standing waves in the intersecting area as ${\bold E}_{1x}=\hat{z}\bar E_{1x}e^{i(\varphi_{1x}+\varphi_{1z}+\varphi_{L})/2}\cos\left(k_0x+\alpha\right)$ and
${\bold E}_{1z}=\hat{x}\bar E_{1z}e^{i(\varphi_{1z}+\varphi_{1x}+\varphi_{L})/2}\cos\left(k_0z+\beta\right)$, where $\bar E_{1x/1z}$ are amplitudes, $\varphi_{1x/1z}$ are the initial phases of the incident lights from $x/z$ directions, and the phase $\varphi_{L}=k_0L$ is acquired through the optical path $L$ from intersecting point to mirror $M_1$, then to $M_2$, and back to the intersecting point, with $k_0=\omega_1/c$ and $\alpha(\beta)=[\varphi_{1x(z)}-\varphi_{1z(x)}-\varphi_{L}]/2$. For alkali atoms, we can show that the optical potential generated by linearly polarized lights is spin-independent when the detuning $\Delta$ is much larger than the hyperfine structure splittings (see Methods). The lattice potential then takes the form
\begin{equation}
V_{\rm latt}(x,z)=V_{0x}\cos^2(k_0x+\alpha)+V_{0z}\cos^2(k_0z+\beta), \label{lattice}
\end{equation}
where $V_{0x/0z}=|\Omega_{x/z}|^2/\Delta$. The Rabi frequency amplitudes of the standing waves $\Omega_{x/z}\equiv{\bf d}_{\rm eff}\cdot{\bf\bar E}_{1x/1z}$, where ${\bf\bar E}_{1x/1z}=\bar E_{1x/1z}\hat z/\hat x$, and the effective dipole matrix ${\bf d}_{\rm eff}$ takes into account the transitions from a ground state ($g_{\uparrow,\downarrow}$) to all relevant excited states in $D_1$ and $D_2$ lines [Fig.~\ref{figure1}{\bf b}]. The lattice potential induces spin-conserved hopping as illustrated in Fig.~\ref{figure1}{\bf d}.

The Raman couplings are induced when another beam $E_{2z}$ of frequency $\omega_2$ is applied and incident from the $z$ direction. The light components $E_{1z}$ and $E_{2z}$ can be generated from the aforementioned single laser source via an acoustic-optic modulator (AOM) which controls their frequency difference $\delta\omega=\omega_2-\omega_1$ and amplitude ratio $\bar E_{2z}/\bar E_{1z}$. The light $E_{2z}$ generates plane-wave fields ${\bold E}_{2z}=\hat{x}\bar E_{2z}e^{i(k_0z+\varphi_2)}$ and
${\bold E}_{2x}=\hat{z}\bar E_{2x}e^{i(-k_0x+\varphi_2+\varphi_L+\delta\varphi_L)}$,
with $\varphi_2$ being its initial phase. The relative phase $\delta\varphi_L=L\delta\omega/c$ acquired by $E_{2x}$ is a crucial parameter, which can be readily manipulated by changing the optical path $L$ or the relative frequency $\delta\omega$, and it controls the dimensionality of the generated SO coupling.
The standing-wave and plane-wave beams form a {\it double-$\Lambda$} type configuration as shown in Fig.~\ref{figure1}{\bf c}, with
$E_{1x}$ and $E_{2z}$ generating one Raman potential via $|F,0\rangle$ in the form
$M_{0x}\cos(k_0x+\alpha)e^{i(k_0z+\beta)+i(\varphi_2-\varphi_{1z})}$,
and $E_{2z}$ and $E_{2x}$ producing another one via $|F,-1\rangle$ as
$M_{0y}\cos(k_0x+\beta)e^{-i(k_0z+\alpha)+i(\varphi_2-\varphi_{1z})+i\delta\varphi_L}$.
Note that for the present blue-detuned lattice, atoms are located in the region of minimum intensity of lattice fields. It follows that terms like $\cos(k_0x+\alpha)\cos(k_0z+\beta)$, which is antisymmetric with respect to each lattice site in both $x$ and $z$ directions, have negligible contribution to the low-band physics. Neglecting such terms, we obtain the Raman coupling potentials by
\begin{eqnarray}
{\cal M}_x(x,z)=M_{x}-M_{y}\cos\delta\varphi_L, \ \quad{\cal M}_{y}(x,z)=-M_y\sin\delta\varphi_L. \label{Raman}
\end{eqnarray}
Here $M_x=M_{0x}\cos(k_0x+\alpha)\sin(k_0z+\beta)$ and $M_{y}=M_{0y}\cos(k_0z+\beta)\sin(k_0x+\alpha)$ with $M_{0x}/M_{0y}=\bar E_{1x}/\bar E_{1z}$ (see Methods). Together with an effective Zeeman term with $m_z=\delta/2$ which is controlled by tuning the two-photon detuning $\delta$ [Fig.~\ref{figure1}{\bf c}], we then reach the effective Hamiltonian~\eqref{Ham}. It is seen that the Raman potential $M_{x/y}$ is antisymmetric with respect to each lattice site along the $\hat{x}/\hat z$ direction [Fig.~\ref{figure1}{\bf e/f}]. This feature has an important consequence that $M_x$ ($M_y$) results in the spin-flipped hopping only along $x (z)$ direction. Moreover, the phase difference $\delta\varphi_L$ governs the relative strength of $\sigma_x$ and $\sigma_y$ terms in the Raman potential, and thus determines the dimensionality of the SO coupling. For example, if setting $\delta\omega=50$MHz, we have $\delta\varphi_L=\pi/2$ for $L=1.5$m, which gives the optimal 2D SO coupling. Further increasing the optical path to $L=3.0$m gives $\delta\varphi_L=\pi$, and the SO coupling becomes 1D form. This enables a fully controllable study on the crossover between 2D and 1D SO couplings by tuning $\delta\varphi_L$, and provides a comparison measurement to confirm the realization of 2D SO interaction.

\begin{figure}[t]
\begin{center}
\ifthenelse{\boolean{SubmittedVersion}}{}{\includegraphics[width= 125 mm]{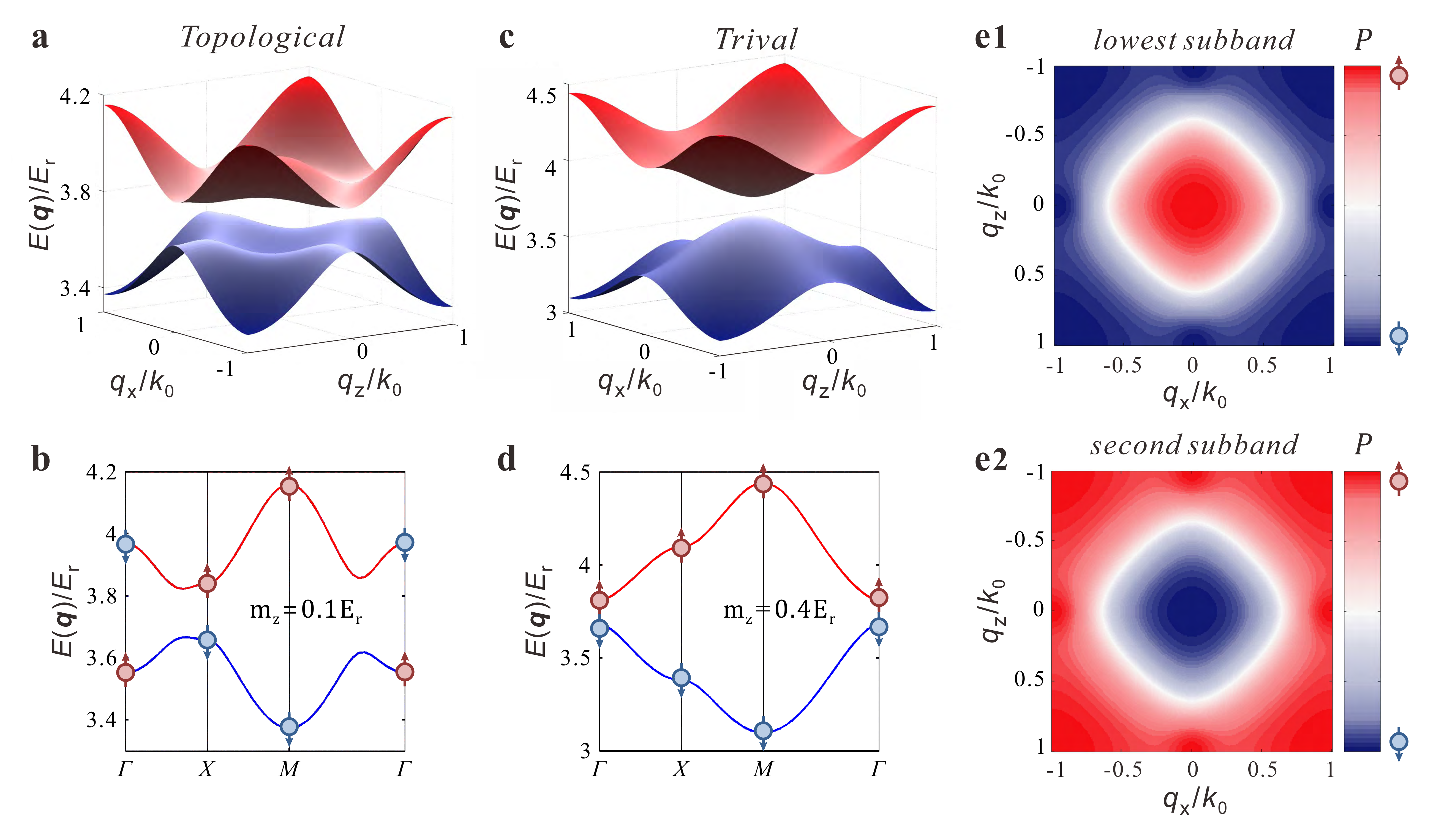}}
\end{center}
\caption{ {\bf Band structure and spin texture with 2D SO interaction.}
An example of gapped band structure with nontrivial band topology ({\bf a}), spin texture along the loop ${\it \Gamma}$-$X$-$M$-${\it \Gamma}$ ({\bf b}), and spin polarization distributions $\langle\sigma_z\rangle$ of the lowest band ({\bf e1}) and second band ({\bf e2}) for $m_z=0.1E_{\rm r}$.
{\bf c-d}, An example of trivial band with gapped band structure ({\bf c}) and spin texture along the loop ${\it \Gamma}$-$X$-$M$-${\it \Gamma}$ ({\bf d}) for $m_z=0.4E_{\rm r}$.
In both cases, we take $V_{0x}=V_{0z}=5E_{\rm r}$, $M_{0x}=M_{0y}=1.2E_{\rm r}$ and $\delta\varphi_L=\pi/2$.
} \label{figure2}
\end{figure}

The Hamiltonian (\ref{Ham}) can lead to quantum anomalous Hall phase~\cite{Liu2014}, and it has an inversion symmetry defined by $(\sigma_z\otimes\hat R_{2D})H(\sigma_z\otimes\hat R_{2D})^{-1}=H$,
where the 2D spatial operator $\hat R_{\rm 2D}$ transforms the Bravais lattice vector ${\bf R}\to-{\bf R}$.
It was shown in Ref.~\cite{Liu2013} that the topology of gapped Bloch bands with such symmetry can be determined
by the product of the spin-polarizations at four highly symmetric momenta $\Theta=\Pi_{j=1}^4{\rm sgn}[P({\bf \Lambda}_j)]$, where $P({\bf\Lambda}_j)$ denotes the spin-polarization and the momenta $\{{\bf \Lambda}_j\}=\{\Gamma(0,0),X_1(0,\pi),X_2(\pi,0),M(\pi,\pi)\}$.
The topological (trivial) phase corresponds to $\Theta=-1$ ($+1$). We examine the topological properties by exactly diagonalizing $H$ and calculating $P({\bf\Lambda}_j)$ at symmetry momenta.
Two typical examples are shown in Fig.~\ref{figure2}, with $V_{0x}=V_{0z}=5E_{\rm r}$, $M_{0x}=M_{0y}=1.2E_{\rm r}$,
$\delta\varphi_L=\pi/2$ and $m_z=0.1E_{\rm r}$ [(Fig.~\ref{figure2}({\bf a,b,e1,e2})] or $m_z=0.4E_{\rm r}$~[Fig.~\ref{figure2}({\bf c,d})], where $E_{\rm r}=\hbar^2k_0^2/2m$ is the recoil energy.
For the chosen parameters, the lowest two subbands are gapped~[(Fig.~\ref{figure2}({\bf a-d})]. When $m_z=0.1E_{\rm r}$,
the spin polarizations at the ${\it \Gamma}$ [$(0,0)$] and $M$ [$(\pi,\pi)$] points are opposite~[Fig.~\ref{figure2}({\bf b,e1,e2})],
which indicates that the band is topologically nontrivial. In contrast, the polarizations are the same for  $m_z=0.4E_{\rm r}$~[Fig.~\ref{figure2}{\bf d}] and the band is trivial.

The present scheme displays several essential advantages.
(i) The fluctuations, e.g. due to the mirror oscillations, have very tiny effect on $L$ and thus cannot affect $\delta\varphi_L$. Also, the initial phases of lights are absorbed into $\alpha,\beta$, which only globally shift the the optical Raman lattice. Thus the present scheme is of topological stability, namely, the realization is intrinsically immune to any phase fluctuations in the setting.
This avoids phase locking, a challenging task in practical realizations.
(ii) As long as $\bar E_{1x}=\bar E_{1z}$ and $\bar E_{2x}=\bar E_{2z}$, which are easily satisfied, the system becomes uniform in $x$ and $z$ directions:
$V_{0x}=V_{0z}$ and  $M_{0x} =M_{0y}$. No fine tuning of optical potentials is requested.
(iii) All the coupling beams can be created from only a single laser source, simplifying the experimental layout.
(iv) More importantly, due to the stability of the double-$\Lambda$ Raman configuration in Fig.~\ref{figure1}{\bf c}, the heating rate,
as determined by the detuning and light field strength, is at the same level for the 1D SO coupling.
These advantages enable that the present scheme is immediately feasible
in ultracold atom experiments within the current technology.

\begin{figure}[t]
\begin{center}
\ifthenelse{\boolean{SubmittedVersion}}{}{\includegraphics[width= 125 mm]{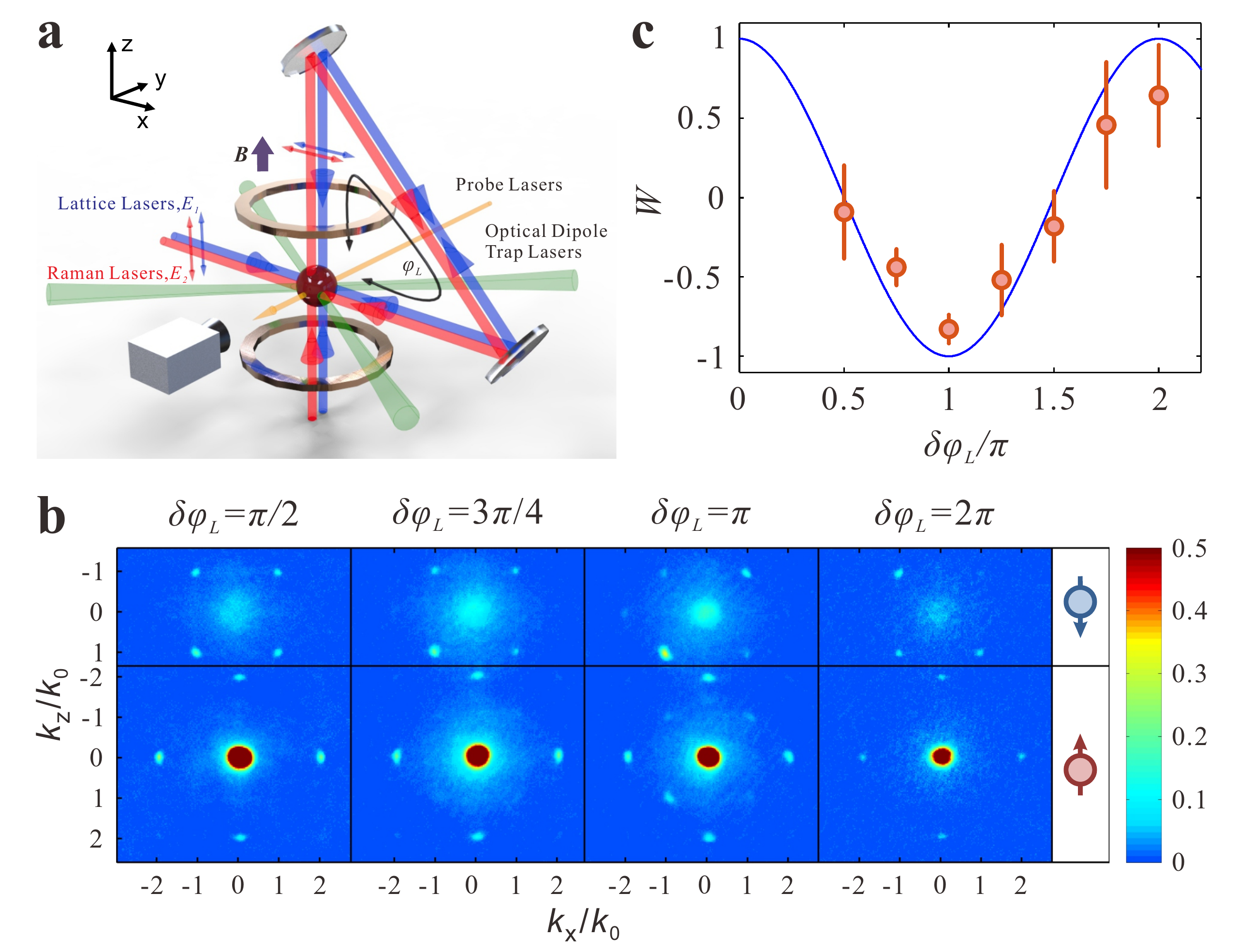}}
\end{center}
\caption{{\bf Experimental realization of 2D SO interaction and 1D-2D crossover.} {\bf a}, Experimental setup. {\bf b}, Spin-resolved TOF images of BEC atoms for
$\delta\varphi_L=\pi/2$, $\delta\varphi_L=3\pi/4$, $\delta\varphi_L=\pi$ and $\delta\varphi_L=2\pi$, respectively. The other parameters are measured as $V_{0x}=V_{0z}=4.16E_{\rm r}$, $M_{0x}=M_{0y}=1.32E_{\rm r}$ and $m_z=0$.
{\bf c}, Measured imbalance $W$ between the Raman coupling induced atoms in the two diagonal directions as a function of the relative phase $\delta\varphi_L$, compared to a cosine curve $\cos\delta\varphi_L$.
The results are averaged over about 30 TOF images. } \label{figure3}
\end{figure}


In our experiment, a BEC of about $1.5\times10^5$ $^{87}$Rb atoms in the
state $|1,-1\rangle$ is prepared in a crossed optical dipole trap with trapping frequencies of $\{\omega_x,\omega_y,\omega_z\}=2\pi\times\{45,45,55\}$Hz.
A bias magnetic field of 49.6G is applied along the $\hat{z}$ direction to generate the Zeeman splitting and determine the quantization axis.
As shown in Fig.~\ref{figure3}{\bf a}, three laser beams in the $x$-$z$ plane with wavelength of 767nm illuminate the atoms for the generation of the Hamiltonian (1).
Among them, a pair of counter-propagating lasers with the same frequency $\omega_1$ (the blue lines in Fig.~\ref{figure3}{\bf a} labelled as ``lattice lasers")
produce the two-dimensional optical lattice. These two lasers are incident along the $\hat{x}$ and $\hat{z}$ directions, respectively,
and reflected by two mirrors $M_1$ and $M_2$ to form the standing waves in both directions.
The polarizations are set in the $x$-$z$ plane so that the interferences between $\hat{x}$ and $\hat{z}$ directions are automatically avoided.
The third laser with frequency $\omega_2$ (the red line in Fig.~\ref{figure3}{\bf a} labelled as ``Raman laser") is a running wave, which is incoming along the $\hat{z}$ direction
with the same polarization as the lattice lasers.
All the  three laser beams are generated from the same Ti:Sapphire laser and the frequencies and amplitudes of these beams are controlled
by two phase-locked AOMs. Thus, the phase coherence is automatically kept, and no additional phase locking is needed.
The Raman and lattice lasers are also coupled into the same optical fiber and then lead to the science chamber,
which helps to avoid the phase noise due to the imperfect overlap in propagation.
The frequency difference $\omega_1-\omega_2$ is set to $35$ MHz to match the Zeeman splitting between $|1,-1\rangle$ and $|1,0\rangle$ states.
The $|1,1\rangle$ state is effectively suppressed due to a large quadratic Zeeman splitting and the system could be treated as a 2-level system.
The detuning $m_z$ can be adjusted by tuning the bias magnetic field.
By controlling the intensities of the lattice and Raman lights, we set the lattice depth $V_{0x}=V_{0z}$ and Raman coupling strength $M_{0x}=M_{0y}$.

In the experiment, the BEC is first prepared in the dipole trap with the bias magnetic field being switched on.
Then, the intensities of the lattice and the Raman beams are simultaneously ramped up to the setting value in 40 ms.
As a consequence, the BEC atoms are adiabatically loaded in the local minimum
of the lowest band at the ${\it \Gamma}$ point (for details, please see Methods).
The phase difference $\delta\varphi_L$ in the Hamiltonian can be achieved
by setting the propagating length between the two mirrors $M_1$ and $M_2$.
The detection is done in the same way as in our previous experiments \cite{Jinyi2012, Cong2014, Cong2015}.
The spin-resolved time-of-flight (TOF) imaging is taken after all the laser beams and the bias magnetic field
are suddenly turned off and a subsequent free expansion for 24ms within a gradient magnetic field to resolve both the momentum and spin.

\begin{figure}[t]
\begin{center}
\ifthenelse{\boolean{SubmittedVersion}}{}{\includegraphics[width= 125 mm]{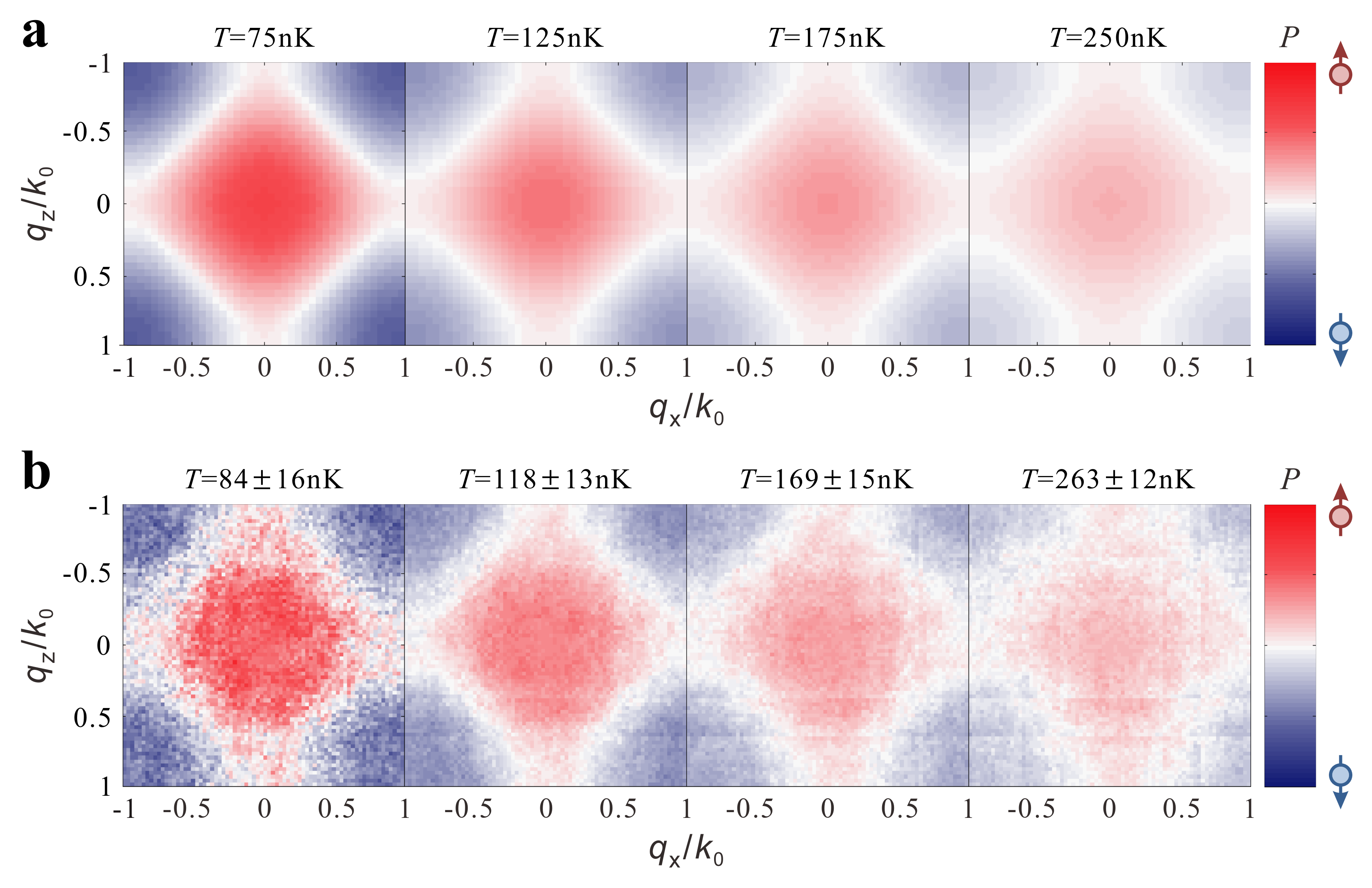}}
\end{center}
\caption{{\bf Spin texture at different temperatures.} {\bf a}, Numerical calculations for the temperatures $T=250$nK, $175$nK, $125$nK, and $75$nK.
{\bf b}, Experimental measurements of spin polarization at different measured temperature. The parameters for the theoretical calculation are the same as the experimental parameters: $V_{0x}=V_{0z}=4.16E_{\rm r}$, $M_{0x}=M_{0y}=1.32E_{\rm r}$ and $m_z=0$.} \label{figure4}
\end{figure}

\begin{figure}[t]
\begin{center}
\ifthenelse{\boolean{SubmittedVersion}}{}{\includegraphics[width= 135 mm]{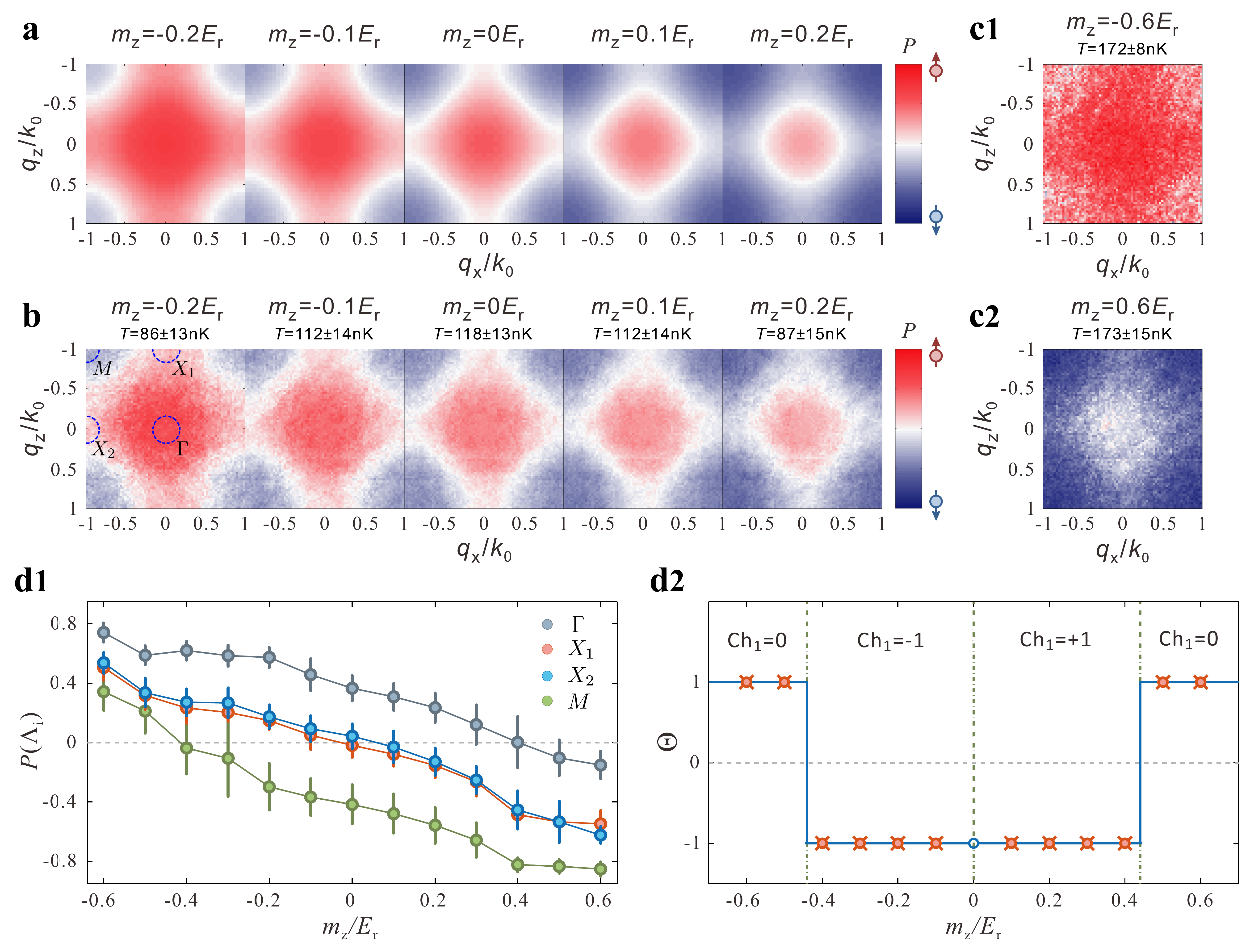}}
\end{center}
\caption{{\bf Spin texture and band topology.} ({\bf a-b}), Spin texture at different $m_z$ by tuning the two-photon detuning.
Experimental measurements ({\bf b}) are compared to numerical calculations at $T=100$nK ({\bf a}).
({\bf c}), Measured spin texture in topologically trivial bands at $m_z=-0.6E_{\rm r}$ ({\bf c1}) and $m_z=0.6E_{\rm r}$ ({\bf c2}).
{\bf d}, Measured spin polarization $P({\bf\Lambda}_j)$ at the four symmetric momenta $\{{\bf\Lambda}_j\}=\{{\it \Gamma},X_1,X_2,M\}$ as a function of $m_z$ ({\bf d1}) and
the product $\Theta=\Pi_{j=1}^4{\rm sgn}[P({\bf \Lambda}_j)]$ ({\bf d2}), which determines the Chern number ${\rm Ch}_1$ and
characterizes the topology of the band.
In all the cases, we set $V_{0x}=V_{0z}=4.16E_{\rm r}$ and $M_{0x}=M_{0y}=1.32E_{\rm r}$.
} \label{figure5}
\end{figure}


To demonstrate the realization of 2D SO coupling, we study the crossover effect in the BEC regime
by tuning the relative phase $\delta\varphi_L$. At $m_z=0$ and by preparing the atoms in the spin-up state, we can adiabatically load the $^{87}$Rb condensate into the ${\it \Gamma}$ point. Then we perform the spin-resolved TOF expansion, which projects Bloch states onto free momentum states with fixed spin polarizations.
Fig.~\ref{figure3}{\bf b} shows the TOF images for various values of $\delta\varphi_L$.
For spin-up ($|\uparrow\rangle$) state, five atom clouds are observed:
besides the major BEC cloud retained at momentum $(k_x,k_z)=(0,0)$, four small fractions of BEC clouds are transferred to momenta $(\pm2k_0,0)$ and $(0,\pm2k_0)$ by the first-order transition due to the lattice potential $V_{\rm latt}$.
The SO coupling is reflected by the two or four small BEC clouds in the state $|\downarrow\rangle$, depending on $\delta\varphi_L$, at the four diagonal corners with momenta $(\pm k_0,\pm k_0)$.
These atom clouds are generated by the Raman potentials, which flip spin and transfer momenta of magnitude $\sqrt{2} k_0$ along the diagonal directions.
As given in Eq.~(\ref{Raman}), the Raman coupling potentials ${\cal M}_x$ and ${\cal M}_y$ depend on $\delta\varphi_L$.
In particular, for $\delta\varphi_L=\pi/2$, we find that four
small clouds in state $|\downarrow\rangle$ with TOF momentum $\vec k=(\pm k_0,\pm k_0)$ are observed [Fig.~\ref{figure3}{\bf b}], implying the
2D SO coupling in the condensate. On the other hand, by tuning the relative phase to $\delta\varphi_L=3\pi/4$, the population of the atom clouds in the two diagonal directions becomes imbalanced. Furthermore, the system reduces to the 1D SO coupling when $\delta\varphi_L= \pi$ and $2 \pi$, with
${\cal M}_x=M_x \pm M_y$ and ${\cal M}_y=0$.
In this case, the Raman pumping only results in a single diagonal pair of BEC clouds, as shown clearly in Fig.~\ref{figure3}{\bf b} with $\delta\varphi_L = \pi$ and $2 \pi$.
This is similar to the 1D SO coupling in the free space in our previous experiments~\cite{Jinyi2012}, where
the Raman coupling flips the atom spin and generates a pair of atom clouds with opposite momenta.
To quantify crossover effect, we define $W=({\cal N}_{\hat{x}-\hat{z}}-{\cal N}_{\hat{x}+\hat{z}})/ ({\cal N}_{\hat{x}-\hat{z}}+{\cal N}_{\hat{x}+\hat{z}})$
to characterize the imbalance of the Raman coupling induced atom clouds,
with ${\cal N}_{\hat{x} \pm \hat{z}}$ denoting the atom number of the two BEC clouds
along the diagonal $\hat{x} \pm \hat{z}$ direction.
The result of $W$ is shown in Fig.~\ref{figure3}{\bf c}, and is characterized well by a simple cosine curve $\cos\delta\varphi_L$, signifying clearly the crossover between 2D and 1D SO couplings realized in the present BEC regime.

Next, we focus on the case of 2D isotropic SO coupling with $\delta\varphi_L=\pi/2$, measure the spin distribution in the first Brillouin zone,
and detect topologically trivial and nontrivial bands by varying the two-photon detuning which governs $m_z$.
For this purpose, we need a cloud of atoms with appropriate temperature such that the lowest band is occupied
by sufficient number of atoms while the population of atoms in the higher bands should still be small.
A similar procedure used in the above BEC measurement is followed except that the atoms are cooled to relatively higher temperatures,
which are measured in a posteriori according to the momentum distribution of hot atoms.
After a TOF expansion, we obtain the atom distributions of both spin-up and spin-down states in the momentum space,
and then map them back to the Bloch momentum space according to the plane-wave expansion of eigenfunctions (see Methods).
We define the spin polarization $P({\bf q})=[n_\uparrow({\bf q})-n_\downarrow({\bf q})]/[n_\uparrow({\bf q})+n_\downarrow({\bf q})]$, with
$n_{\uparrow,\downarrow}({\bf q})$ being the density of atoms of the corresponding spin state in the first Brillouin zone.
Figure~\ref{figure4} (a) and (b) show the numerical results and experimentally measured spin polarizations at different temperatures, respectively, for $m_z=0$, $V_{0x}=V_{0z}=4.16E_{\rm r}$ and $M_{0x}=M_{0y}=1.32E_{\rm r}$. In doing numerical simulation the finite temperature effect is taken into account based on the Bose-Einstein statistics
$f(E)=1/[e^{(E_{\bf q}-\mu)/k_{\rm B}T}-1]$, with $k_{\rm B}$ the Boltzmann constant and $E_{\bf q}$
given by band energies of the Hamiltonian (\ref{Ham}), plus the kinetic energy $\hbar^2q_y^2/2m$ due to the motion in the out-of-lattice plane ($y$) direction. The average atom density is taken as $n=3\times10^{19}m^{-3}$ , which determines the chemical potential $\mu$. It can be seen that the theoretical and experimental results agree well, which demonstrates the feasibility and the reliability of the spin polarization measurement.
Furthermore, the results in Fig.~\ref{figure4} suggest that a temperature around $T=100$nK is preferred in our experiment to extract the spin texture information of the lowest band. In comparison,
if the temperature is too high, atoms are distributed over several bands and the visibility of the spin polarization will be greatly reduced,
while a too low temperature can also reduce the experimental resolution since the atoms will be mostly condensed at the band bottom.

We then measure the spin polarization as a function of detuning $m_z$ to reveal the topology of the lowest energy band, with $V_{0x}=V_{0z}=4.16E_{\rm r}$ and $M_{0x}=M_{0y}=1.32E_{\rm r}$.
The numerical calculations and TOF measured images of $P({\bf q})$ are given in Fig.~\ref{figure5}{\bf a} and ({\bf b,c}), respectively, which also show agreement between the theoretical and experimental results.
In Fig.~\ref{figure5}{\bf d}, we plot the values of polarization $P({\bf \Lambda}_j)$ for the four highly symmetric momenta ${\it \Gamma}$, $X_1$, $M$ and $X_2$ (d1).
It can be seen that $P(X_1)$ and $P(X_2)$ always have the same sign, while the signs of $P({\it \Gamma})$ and $P(M)$
are opposite for small $|m_z|$ and the same for large $|m_z|$, with a transition occurring at the critical value of $|m_z^c|$ which is a bit larger than $0.4E_r$.
From the measured spin polarizations, the product $\Theta$ and the corresponding Chern number, given by ${\rm Ch_1}=-\frac{1-\Theta}{4}\sum_{j=1}^4{\rm sgn}[P({\bf \Lambda}_j)]$ according to Ref.~\cite{Liu2013}, are readily read off and plotted in Fig.~\ref{figure5} ({\bf d1,d2}).
The results agree well with numerical calculations which predict two transition points between the topologically trivial and nontrivial bands
near $ m^c_z = \pm 0.44E_{\rm r}$.
This confirms that for the 2D SO coupled system realized in the present experiment, the energy band is  topologically nontrivial when $0<|m_z|<|m_z^c|$, while it is trivial for $|m_z|>|m_z^c|$.

\noindent{\bf Discussion}

With the advantages of topological stability and small heating rate, the present experimental realization of 2D SO interaction opens a broad avenue in exploring novel quantum physics, ranging from quantum spin dynamics to exotic topological phases. For example, by applying an external force in the $x$-$z$ plane (e.g. the gravity force in the $-z$ direction), the Bloch oscillation with a nontrivial Hall effect can be obtained. By properly adjusting system parameters, the Berry phase effect, $\bold k$-space monopole, and Landau-Zener effect across the topological transition points can be probed with the current technology. On the other hand, thanks to the high controllability of the present realization, the SO interaction can be readily switched on and off, and be adjusted between 1D and 2D limits. This may lead to rich quench spin dynamics in the optical lattice with nontrivial band structure. It is also noteworthy that, with the present SO coupling in optical lattice, one may explore completely novel states of matter such as SO coupled Mott insulators with interacting bosons, which have no analogue in solids~\cite{Magneticphase2,Trivedi2012}.

Furthermore, the present optical Raman lattice scheme is generic and can be immediately applied to fermion systems (e.g. $^{40}$K), in which case, the quantum anomalous Hall effect in the single-particle regime and topological superfluid~\cite{Liu2014} or novel magnetic phases~\cite{Magneticphase1} in the interacting regimes will be particularly interesting and important. We emphasize that the topological superfluid is a highly-sought-after phase, for it hosts Majorana quasiparticles which obey non-Abelian statistics~\cite{Ivanov2001,Alicea2011} and have attracted great attention in both condensed matter physics and cold atoms~\cite{Wilczek2009,Franz2013}. The successful realization of 2D SO interaction in our experiment advances an essential step toward the observation of this remarkable state. Finally, while the present study is focused on a 2D lattice system, generalizing our scheme to 3D optical lattices may lead to the realization of topological phases in 3D system, including the Weyl topological semimetals~\cite{Wan2011,Xu2011,Burkov2011,Xu2015,Weng2015}.\\

\bibliographystyle{naturemag}
\bibliography{cvppapers}

\begin{addendum}
 \item We thank Jason Ho, Ting-Fung Jeffrey Poon, and Gyu-Boong Jo for helpful discussions. This work has been supported by the NNSF of China, the CAS, the National Fundamental Research Program (under Grants No. 2013CB922001), and Peking University Initiative
Scientific Research Program. X.J.L. is also support by the Thousand-Young-Talent Program of China.
 \item[Competing Interests] The authors declare that they have no
competing financial interests.
 \item[Author Contributions] Y.J.D., S.C., X.J.L. and J.W. P. conceived the project. X.J.L. proposed the theoretical proposal and performed major calculations. S.C. and J.W.P. designed and supervised the experimental realization. Z.W., W.S., X.T.X. and S.C.J. performed the experiment. L.Z. and B.Z.W. performed numerical calculations. All authors contributed to writing of the manuscript.
 \item[Author Information] Correspondence and requests for materials should be addressed to S. Chen (shuai@ustc.edu.cn), X.-J. Liu (xiongjunliu@pku.edu.cn), or J.-W. Pan (pan@ustc.edu.cn).
\end{addendum}

\ifthenelse{\boolean{SubmittedVersion}}{\processdelayedfloats}{\cleardoublepage}

\begin{methods}
\subsection{Lattice and Raman potentials}
The standing waves forming the square lattice potentials are given by
${\bf E}_{1x}=\hat{z}\bar E_{1x}\exp[i(\varphi_{1x}+\varphi_{1z}+\varphi_{L})/2]\cos(k_0x+\alpha)$ and
${\bf E}_{1z}=\hat{x}\bar E_{1z}\exp[i(\varphi_{1z}+\varphi_{1x}+\varphi_{L})/2]\cos(k_0x+\beta)$.
By a straightforward analysis on both $D_1$ and $D_2$ transitions~[see Fig.~\ref{figure1}{\bf b}], one can show that the lattice
potentials for the two spin states are given by ($\sigma=\uparrow,\downarrow$)
\begin{eqnarray}
V_\sigma&=&\sum_{F}\frac{1}{\Delta}\left(\left|{\Omega}_{\sigma F,1x}^{(3/2)}\right|^2+\left|{\Omega}_{\sigma F,1z}^{(3/2)}\right|^2\right)\nonumber\\
&+&\sum_{F}\frac{1}{\Delta+\Delta_s}\left(\left|{\Omega}_{\sigma F,1x}^{(1/2)}\right|^2+\left|{\Omega}_{\sigma F,1z}^{(1/2)}\right|^2\right),
\end{eqnarray}
where the Rabi frequencies ${\Omega}_{\uparrow F,1x}^{(J)}=\langle\uparrow|ez|F,-1,J\rangle \hat e_z\cdot{\bf E}_{1x}$ and
$\Omega_{\uparrow F,1z}^{(J)}=\langle\uparrow|ex|F,0,J\rangle\hat e_x\cdot {\bf E}_{1z}+\langle\uparrow|ex|F,-2,J\rangle\hat e_x\cdot {\bf E}_{1z}$ with $e$ being the charge of an electron, the angular momentum index represents the $D_1$ (for $J=1/2$) and $D_2$ (for $J=3/2$) transitions, and $|F,m_F,J\rangle$ denoting the excited atomic states. The Rabi frequencies ${\Omega}_{\downarrow F,1x(1z)}^{(J)}$ for spin-down state
take similar forms as ${\Omega}_{\uparrow F,1x(1z)}^{(J)}$ by replacing $m_F$ with $m_F+1$. From the data of dipole matrix elements
of $^{87}$Rb one can show that
\begin{equation}
V_{\uparrow}=V_{\downarrow}=\frac{1}{3}\left(\frac{\alpha_{D_2}^2}{\Delta}+\frac{\alpha_{D_1}^2}{\Delta+\Delta_s}\right)(|{\bf E}_{1x}|^2+|{\bf E}_{1z}|^2),
\end{equation}
where $\alpha_{D_2}\equiv|\langle J=1/2||e{r}||J'=3/2\rangle|$, $\alpha_{D_1}\equiv|\langle J=1/2||e{r}||J'=1/2\rangle|$, and
$\alpha_{D_2}=\sqrt{2}\alpha_{D_1}=4.227ea_0$~\cite{Long2015} with $a_0$ the Bohr radius.
Thus the lattice potential takes the form as shown in Eq.~\eqref{lattice} and the amplitudes $V_{0x(0z)}=\frac{3\Delta+2\Delta_s}{3\Delta(\Delta+\Delta_s)}\alpha_{D_1}^2\bar E_{1x(1z)}^2$ are spin-independent.

The two Raman coupling potentials are generated by the light components ${\bf E}_{1x}$, ${\bf E}_{2z}$ and ${\bf E}_{1z}$, ${\bf E}_{2x}$, respectively [Fig.~\ref{figure1}{\bf c}]. Similar as the calculation of lattice potentials, we take into account the contributions from both $D_1$ and $D_2$ transitions, and obtain the Raman coupling amplitudes by $M_{0x(0y)}=\frac{1}{6\sqrt{2}}\bigr(\frac{1}{\Delta}-\frac{1}{\Delta+\Delta_s}\bigr)\alpha_{D_1}^2\bar E_{1x(1z)}\bar E_{2z(2x)}$~\cite{Long2015}.
From Fig.~\ref{figure1}{\bf a}, we can naturally set that the amplitudes satisfy $\bar E_{1x}=\bar E_{1z}$ and $\bar E_{2x}=\bar E_{2z}$. It then follows that $V_{0x}=V_{0z}$ and $M_{0x}=M_{0y}$, namely, the lattice potential and Raman couplings are uniform in $x$ and $z$ directions, which greatly facilitates the manipulation of the created optical Raman lattice.

\subsection{Exact diagonalization}
The Hamiltonian~\eqref{Ham} can be exactly diagonalized with plane-wave expansions. For this we construct a complete set basis of plane waves $\{\psi_{m,n}^{\uparrow}(\bold k), \psi_{p,q}^{\downarrow}(\bold k)\}$ with the spinors
\begin{eqnarray}\label{basis}
\psi_{m,n}^{\uparrow}(\bold q)=\frac{1}{\sqrt{S}}e^{i(q_x+2k_0m)x}e^{i(q_z+2k_0n)z}, \ \
\psi_{p,l}^{\downarrow}(\bold q)=\frac{1}{\sqrt{S}}e^{i(q_x+2k_0p+k_0)x}e^{i(q_z+2k_0l+k_0)z},
\end{eqnarray}
where $m$, $n$, $p$ and $l$ are integers, and $S$ denotes the system area.
The eigenfunctions for the Hamiltonian (\ref{Ham}) with eigenvalue $E_{\bold q}$ can be expressed as
\begin{equation}\label{efunction}
|\Psi_{E_\bold q}\rangle=\sum_{m,n}a_{m,n}\psi_{m,n}^\uparrow(\bold q)|\uparrow\rangle+\sum_{p,l}b_{p,l}\psi_{p,l}^\downarrow(\bold q)|\downarrow\rangle,
\end{equation}
with $a_{m,n}$ and $b_{p,l}$ the coefficients.
Using the relation $\langle\psi_{m',n'}^{\sigma}|\psi_{m,n}^{\sigma'}\rangle=\delta_{m',m}\delta_{n',n}\delta_{\sigma\sigma'}$ one can write
the Hamiltonian (\ref{Ham}) in the matrix form. To diagonalize the Hamiltonian one needs to choose a proper cut-off $M$ for the expansion in Eq.~\eqref{efunction}, so that $|m|,|n|,|p|,|l|\leq M$. We have verified that for our experimental parameter regime, the Hamiltonian is diagonalized with a very high accuracy when $M\geq10$. In this way, we obtain the energy bands $E({\bf q})$ and the corresponding eigenfunctions (\ref{efunction}). From~\eqref{basis} and~\eqref{efunction} one can also obtain a mapping between the Bloch eigenstates and TOF states.

\subsection{State preparation and Detection}
The $^{87}$Rb atoms in $|1,-1\rangle$ state are
trapped and cooled in a crossed optical dipole trap as in Refs.\cite{Jinyi2012, Cong2014, Cong2015}.
A bias magnetic field along the $\hat{z}$ direction with $B=49.6$G is ramped up in the final stage of the evaporation, 4 seconds before the condensation.
Here the quadratic Zeemen shift $\epsilon=353.5$ kHz, which is $90.6$ times of the recoil energy $E_{\rm r}=\hbar^{2}k_{0}^{2}/2m=2\pi\times3.9$kHz for 767 nm lasers
and thus effectively suppresses the $|1,1\rangle$ state for Raman coupling.
The lattice and Raman lasers with wavelength of $\lambda=767$nm illuminate the atoms along $\hat{x}$ and $\hat{z}$ directions (as shown in Fig.~\ref{figure3}{\bf a}).
They are ramped up to the setting value in 40ms to make sure the loading process is adiabatic and the BEC atoms are prepared at the ${\it \Gamma}$ point
of the lowest band. For the filling of thermal atoms to detect the spin distribution, the intensity of the optical dipole trap laser is controlled in the final evaporation stage to vary the trap depth, which determines the final temperature of the atom cloud. All the other steps are the same as the BEC case.

For detection, the spin-resolved TOF imaging is taken after suddenly switching off all the lasers within $1$ $\rm{\mu s}$ and
a free expansion for 24ms.
During the expansion, 
a gradient magnetic field along $z$-axis is applied to
separate $|1,-1\rangle$ and $|1,0\rangle$ states.

\subsection{Estimation of heating}
The heating is caused by all the lasers including the dipole trap lights, the lattice and the Raman lights.
The heating rate of the dipole trap is measured to be 18 nK/s, mainly due to the photon scattering and the intensity noise.
It gives the BEC life time of about 10 seconds. The heating rate of the lattice lasers is at the same level with that of the dipole trap laser for $V_0=4.16E_{\rm{r}}$,
The heating rate of the Raman light is estimated to be an order of magnitude higher than the dipole trap at present parameter of $M_0=1.32E_{\rm{r}}$. In the current experiment all these lasers together affect the life time of the BEC with 2D SO coupling to be about 300ms.
The total heating rate is comparable to that of the previous 1D spin-orbit coupling experiment {\cite {Cong2014}}, and the life time is also estimated to be enough
to explore more exotic physics in this system.

\end{methods}

\end{document}